\newcommand\g{\gamma}
\renewcommand\d{\delta}
\newcommand\e{\epsilon}

\newcommand\s{\sigma}

\newcommand\f{\phi}

\renewcommand\o{\omega}

\newcommand\ra{\rightarrow}

\newcommand{\lan}{\langle}
\newcommand{\ran}{\rangle}

\newcommand{\diracslash}[1]{#1\llap{/\kern2pt}}

\newcommand{\be}{\begin{equation}}
\newcommand{\ee}{\end{equation}}
\newcommand{\bea}{\begin{eqnarray}}
\newcommand{\eea}{\end{eqnarray}}
\newcommand{\ba}[1]{\begin{array}{#1}}
\newcommand{\ea}{\end{array}}
\newcommand{\bep}{\begin{pmatrix}}
\newcommand{\eep}{\end{pmatrix}}

\newcommand{\bt}{\begin{tabular}}
\newcommand{\et}{\end{tabular}}

\newcommand{\beas}{\begin{eqnarray*}}
\newcommand{\eeas}{\end{eqnarray*}}

\documentclass[preprint,prd,aps,floats,nofootinbib,floatfix]{revtex4}
\usepackage{graphicx}
\usepackage[intlimits]{amsmath}
\usepackage[utf8]{inputenc}
\usepackage[T1]{fontenc}
\usepackage{lineno}
\usepackage{textcomp}
\usepackage{subfig}
\usepackage{multirow}
\usepackage{xcolor} 
\usepackage{comment}
\usepackage[a4paper, total={160mm, 240mm}, top={0.5in}]{geometry}
\usepackage{tabularray}
\usepackage{ulem} 
\begin{document}
\title{Open Strange and Open Heavy Flavor mesons in Asymmetric Nuclear Matter within Quark Meson Coupling model }
\author{Arpita Mondal}
\email{arpita.mondal@physics.iitd.ac.in
}
\affiliation{Department of Physics, Indian Institute of Technology, Delhi, Hauz Khas, New Delhi -- 110 016, India}

\author{Amruta Mishra}
\email{amruta@physics.iitd.ac.in}
\affiliation{Department of Physics, Indian Institute of Technology, Delhi, Hauz Khas, New Delhi -- 110 016, India} 
\begin{abstract}
The in-medium properties of open strange ($K$, $\bar{K}$), open charm ($D$, $\bar{D}$), and open bottom ($B$, $\bar{B}$)  mesons are investigated in asymmetric nuclear matter using Quark Meson Coupling (QMC) model. A direct coupling of scalar ($\sigma$, $\delta$) and vector ($\omega$, $\rho$) mesons to the light quarks and anti-quarks of these mesons give rise to the in-medium modification of the properties of the corresponding meson within the model. 
The inclusion of the $\d$ (scalar iso-vector) meson breaks the isospin symmetry for the masses of the light quark and antiquark doublets, causing mass splitting between ($u,\;d$) as well as ($\bar{d},\;\bar{u}$). Consequently, the considered mesons exhibit mass splittings within the isodoublets of $K$, $\bar{K}$, $D$, $\bar{D}$, $B$ and $\bar{B}$ mesons when embedded in asymmetric nuclear matter. In the current study, the interactions of the pseudoscalar meson with the scalar, as well as vector mesons, are considered, which lead to significant medium modifications of the excitation energies of the open strange (charm and bottom) mesons. In asymmetric nuclear matter, due to the interaction of the pseudoscalar meson with the vector iso-vector $\rho$ meson, there is a splitting in the excitation energies of the mesons within the isospin doublets. The isospin effects are seen to be large for high baryon densities.
This study can have significant observable consequences, such as in the production ratios, e.g., $K^+/K^0$, $K^-/\bar{K}^0$, $D^+/D^0$, $D^-/\bar{D}^0$, $B^+/B^0$ and $B^-/\bar{B}^0$ in the upcoming heavy ion collision experiments at FAIR project at GSI, where the experiments are planned to be performed using neutron-rich beams to study the compressed baryonic matter.
\end{abstract}
\maketitle

\def\bfm#1{\mbox{\boldmath $#1$}}
\def\bfs#1{\mbox{\bf #1}}
\section{Introduction}
\label{1}
One of the important research areas in strong interaction physics is the study of in-medium properties of hadrons, as it has direct relevance for Heavy Ion Collision experiments (HIC) as well as for the study of astrophysical compact objects such as neutron stars.
In HIC experiments, two heavy ion beams at ultra-relativistic energies collide and can lead to the liberation of the fundamental constituents of the nucleons forming a deconfined state of quarks and gluons, known as quark gluon plasma (QGP), which cools further via rapid expansion through various stages of evolution. The different probes to study the strongly interacting matter formed in HIC experiments are particle yields, particle spectra as well as their collective flows. These observables provide insights into the medium modifications of hadrons' properties.
Several experiments have indicated that the properties of hadrons undergo medium modifications within the created strongly interacting matter.
For instance, the obtained dilepton spectra at the SPS (Super Proton Synchrotron) \cite{ceres, helios} indicated the medium modifications of vector mesons \cite{Brat, Cass, Mishra1, Mishra2}. A mass reduction of around $3.4\%$ and an increase of the decay width of the $\phi$ meson by a factor of $3.6$ at normal nuclear matter density, $\rho_0$ as compared to the vacuum value of $4.25\; {\rm{MeV}}$ \cite{pdg} were also reported by KEK-PS E325 collaboration \cite{kek}. However, experiments at Laser Electron Photon (LEP) facility at SPring-8 \cite{leps} and Continuous electron beam accelerator facility (CEBAF) large acceptance spectrometer, CLAS, at JLab \cite{clas} presented a broadening of $\f$ meson decay width without any mass shift, which leaves the in-medium masses of vector mesons open for further exploration.
For instance, the J-PARC E16 \cite{e16} and E29 \cite{e29} collaborations are actively investigating mass shifts of the $\phi$ meson. 
On the other hand, as per the Particle Data Group (PDG) \cite{pdg}, the branching ratio of $\f \ra K \bar{K}$ ($K^+K^-$ and $K^0\bar{K}^0$) is approximately 83\%, which is greater than all other decay channels of the $\f$ meson, suggesting the modifications of the $\f$ meson properties through the modification of kaons and antikaons in the nuclear matter.
In 1992, Ko \textit{et al.} \cite{ko} reported a mass shift of approximately $\sim 2\%$ and a decay width of $25\; \rm{MeV}$ for $\f$ meson at normal density ($\rho_0$) considering $K-\bar{K}$ loop using chiral perturbation theory. Using the QCD sum rule approach, Hatsuda \textit{et al.} \cite{hatsuda} obtained a $1.5\%-3\%$ mass shift of $\f$ meson at $\rho_0$ through the medium modification of $K$ meson, for more detail, see e.g., \cite{martinez, martinez1}. Mishra \textit{et al.} \cite{Mishra3} showed a substantial enhancement of the decay width of $\f$ meson in the medium using a field
theoretic model of composite hadrons with quark (and antiquark) constituents considering the in-medium properties of kaons. Overall, both the experimental \cite{fopi,kaos1,kaos2,kaos3,kaos4} and theoretical studies \cite{MishraG, MishraS, MishraK, sarmistha, mazumdar, kumar10, t1,t2,t3,t4,t5,t6,t7,t8,t9,t10} on $K$ mesons have shown that the in-medium properties of the kaons influence the yields as well as the collective flow patterns of the kaons.

The open heavy flavor mesons ($D$ and $B$) exhibit behavior analogous to the $K$ mesons in the nuclear matter, as they share similar constituents with the strange quark replaced by heavy (charm or bottom) quarks.
Various analyses of dilepton spectra by NA50 \cite{na50} at CERN SPS, PHENIX \cite{phenix} at relativistic heavy ion collider, RHIC, and E705 \cite{e705} at Fermilab, indicated the medium modifications of charm mesons. Recently the ALICE collaboration at LHCb \cite{alice} has reported the first study on $\bar{D}N$ interaction, where using $pp$ collision data obtained during the LHC run 2, an attractive $pD^-$ interaction has been predicted.
Besides, there exist several works \cite{111, 112, hayashigaki, lee, tolos, kumar1, spmishra, ghosh, mishra5, souradeep}, using different theoretical models, which demonstrated that due to the medium modifications of $D$ mesons, the decays of excited charmonium states to $D\bar{D}$ pair can result in open charm enhancement in a hadronic environment. 
Likewise, the properties of open bottom mesons can undergo modifications in hadronic matter, and there are several studies \cite{hilger,rahul, yamaguchi, DP, ND, AMSPM} exploring their behavior using various approaches. Since heavy flavored mesons are regarded as valuable tools for examining the in-medium properties, especially the open bottom mesons as they do not thermalize at the standard collider energies \cite{b1,b2}, it is suggested that the open bottom mesons are better-suited probes than the other heavy flavor hadrons. Few studies \cite{b1,b2,b3,b4} have investigated the utility of open bottom mesons as probes and their transport properties within hadronic medium.

Among the above studies, most of the theoretical models are based on the hadronic degrees of freedom, such as chiral effective model, coupled channel approach, and others, which are able to explain the in-medium properties of hadrons successfully.
However, to understand the Deep inelastic scattering (DIS) experiments at a high momentum transfer region, it is necessary to consider quark degrees of freedom. Also in 1983, a study by the European Muon Collaboration (EMC) \cite{emc} at JLab, indicated that the internal structure of hadrons plays important consequences on the properties of hadrons in the nuclear medium. Following such aspects, P.A.M. Guichon \cite{guichon}, in 1988, proposed the Quark Meson Coupling (QMC) model, where nucleons are no longer treated as point particles, and the degrees of freedom were considered to be quarks. In the literature \cite{guichon}, nuclear matter properties are described by considering quarks inside the non-overlapping nucleon bags, which interact through the scalar and vector meson fields. In 1994, Saito \textit{et al.} \cite{saito1}, showed that the QMC model quantitatively explains the EMC effects. Afterwards, the QMC model has been very successful in explaining the finite nuclei properties as well as nuclear matter properties \cite{krein1, saito2}. In 1998, Tsushima \textit{et al.} \cite{tsushima1} studied the $KN$ interactions in asymmetric nuclear matter, within this model. Due to the isoscalar ($\o$) vectorial interactions, there is a rise(drop) in the mass of $K(\bar{K})$ meson, whereas, in the isospin asymmetric nuclear matter, the isovector ($\rho$) gives splittings in the excitation energies within the isodoublets of $K$ and $\bar{K}$ mesons. Recently, Mart\'inez \textit{et al.} \cite{martinez} calculated a mass drop of around $13\%$ for kaons within the QMC model, leading to around $1.8\%$ mass drop for $\f$ meson (through the $K-\bar{K}$ loop) at the nuclear matter saturation density. Using the same model, Sibirtsev \textit{et al.} \cite{sibirtsev} studied $D^\pm N$ scattering as well as the $J/\psi$ absorption cross-section through the modifications of $D$ and $\bar{D}$ mesons in the nuclear medium. Recently, in Ref. \cite{bottom}, bottom meson properties are investigated using QMC model. Now, given the isospin imbalance present during the HIC experiments, it is important to investigate how the isospin asymmetry influences the characteristics of the hadrons produced as a consequence of these collisions. 
In 1994, Saito \textit{et al.} \cite{Saito1994} studied such isospin symmetry breaking in nuclear matter by incorporating scalar-isovector ($\d$) meson within the QMC model and also addressed the Nolen-Schiffer anomaly for finite nuclei. Following that work,
Niu \textit{et al.} \cite{niu} investigated neutron star properties in asymmetric nuclear matter and
Santos \textit{et al.} \cite{santos} investigated the saturation properties of the nuclear matter within the QMC$\d$ model.
In the present work, our prime interest is to study the role of scalar-isovector meson field ${\d}$ using the Quark Meson Coupling (QMC) model, following the approach of previous works \cite{Saito1994,niu,liu,santos}, along with the isoscalars ($\s$, $\o$) and isovector (${\rho}$) fields to study the $K(\equiv K^+, K^0),\; \bar{K}(\equiv K^-, \bar{K}^0),\; D(\equiv D^+, D^0),\; \bar{D} (D^-, \bar{D}^0)\; {\rm {and}}\; B(\equiv B^+, B^0 ),\; \bar{B} (\equiv B^-, \bar{B}^0)$ meson properties in the asymmetric nuclear matter (ANM).
The presence of $\delta$ (scalar iso-vector) meson breaks the isospin degeneracy of the masses of the light quark and antiquark doublets and causes mass splitting between ($u,\;d$) as well as ($\bar{d},\;\bar{u}$), which thus causes the mass splittings within the isodoublets of $K$, $\bar{K}$, $D$, $\bar{D}$, $B$ and $\bar{B}$ mesons when embedded in ANM. 
Further, we study the excitation energies of these pseudoscalar mesons considering their interactions with the scalar and vector mesons in asymmetric nuclear matter.
This study could yield observational consequences, e.g., on the particle production ratios of the mesons within the isospin doublets in the upcoming heavy ion collision experiments using neutron-rich beams at ${\rm{\bar{P}ANDA}}$ and compressed baryonic matter (CBM) experiments at the future facility for Antiproton and Ion Research, FAIR (GSI) \cite{panda1, fair, cbm1, cbm2, panda2}, PHENIX collaboration at RHIC at Brookhaven National Laboratory (BNL)\cite{phenix2}, LHCb collaboration at CERN \cite{lhcb}, J-PARC \cite{jparc}, and NA61 Collaboration at CERN-SPS \cite{na61}.
 
The paper is organized as follows:  To begin with, in section \ref{2}, we briefly describe the QMC model, which describes the interactions of the quarks through the meson fields. Then we study the open strange and open heavy flavor mesons properties using this model in section \ref{3}. The model parameters, results, and the discussions are presented in section \ref{4},  where we describe the in-medium behavior of the fields, in-medium masses, and the in-medium excitation energies of these mesons. Finally, in section \ref{s3}, we summarise our findings of the current study.
\section{Quark Meson Coupling (QMC) Model }
\label{2}
In the present study, we investigate the in-medium properties of open strange and open heavy flavor pseudoscalar ($J^P=0^-$) mesons, using the quark meson coupling (QMC) model. Within this model, hadrons are regarded as static spherical MIT bags and the quarks inside the bags interact by exchange of meson fields. 
The direct couplings of scalar-isoscalar $\s$, vector-isoscalar $\o$, scalar-isovector $\d$, and vector-isovector $\rho$ meson fields to the light quark ($u,d$) and light antiquark ($\bar{u},\bar{d}$) constituents
of the hadrons are considered in this framework. In particular, the $\d$- and $\rho$- mesons introduce the isovectorial potentials to simulate the isospin asymmetry within the model.
The couplings of the light quarks and antiquarks with the meson fields modify their properties (masses and energies) in the nuclear matter.
In QMC model, the effective Lagrangian density describing the interactions in asymmetric nuclear matter (ANM) is,
\begin{align}
    \label{A1}
    \mathcal{L}  &=  \bar{\psi}\Big[i\gamma.\partial - \Big(M_N - \tilde{g}_\s({\s}) {\s} - \tilde{g}_\d({\d})\frac{{\tau}^a}{2}{\d}^a\Big) - \gamma^\mu\Big(g_\o {\o}_\mu + g_\rho\frac{{\tau}^a}{2}{{\rho}}_{\mu} ^a\Big)\Big]\psi
\nonumber\\&+\frac{1}{2} ( \partial_\mu {\sigma}\partial^\mu{\sigma} - 
m_\sigma^2{\sigma}^2  ) + \frac{1}{2}(\partial_\mu{{\delta}}^a \partial^\mu {{{\delta}}^a}- m_\delta^2 {{{\delta}}^a}{{{\delta}}^a}) - \left[\frac{1}{4}\o_{\mu\nu} \o^{\mu\nu} - \frac{1}{2}m_\omega^2 {\omega}_\mu {\omega}^\mu\right]\nonumber\\ &- (\frac{1}{4}{{\rho}}_{\mu\nu} ^a{{\rho}}^{\mu\nu, a} - \frac{1}{2}m_\rho^2 {{\rho}}_{\mu}^a{{{\rho}}^{\mu,a})},\quad\quad 
\end{align}
\begin{align}
\hspace{-6.8cm}\rm{where,}\hspace{1.5cm}
\psi = \left(\begin{array}{cc}
    \psi_p       \\
    \psi_n       
\end{array}\right),\;\;\;\;\;
M_N = \begin{pmatrix}
     M_p & 0 \\
    0 &  M_n
    \end{pmatrix};
\end{align} 
$m_\s$, $m_\o$, $m_\d$, and $m_\rho$ are respectively their masses and $\tau^a/2$ is the isospin operator for the nucleons with $a={1,2,3}$. In eq. (\ref{A1}), $\omega_{\mu\nu} = \partial_{\mu}{\omega}_{\nu} - \partial_{\nu}{\omega}_{\mu}$ and $\rho_{\mu\nu} ^a = \partial_{\mu}{{\rho}}_{\nu} ^a - \partial_{\nu}{{\rho}}_{\mu} ^a $ are the field strength tensors corresponding to the vector meson fields $\omega$ and $\rho$, respectively.
The coupling strengths $\tilde{g}_\s({\s})$ and $\tilde{g}_\d({\d})$ correspond to the nonlinear scalar-meson-nucleon interactions and the vector-meson-nucleon couplings are denoted by $g_\o$ and $g_\rho$. These meson fields are treated as classical fields within the mean-field approximation (MFA), which amounts to
 \bea
\label{A4}
{\s} &\to& \lan{\s}\ran \equiv \s, \;\;\;\;\;\;\;\;\;\;\;\;\;\;\;\;\;\;\;\ {{{\d}}^a} \to \lan{{{\d}}^a}\ran \equiv \d^{a3}\d^a = \d^3,\nonumber\\
{\o}_\mu &\to&\lan{\o}_\mu\ran \equiv \d^{\mu 0}\, \o_\mu,\;\;\;\;\;\;\;\;\;\;\;{{\rho}}_{\mu}^a \to \lan{\rho}_{\mu}^a\ran \equiv \d^{\mu 0} \d^{a 3} \rho_{\mu}^a = \d^{\mu 0} \rho_{\mu}^3.
 \eea
In this study, we consider the 
nuclear matter at rest, therefore only $\omega_0$ and $\rho_{0}^3$ components of $\o_\mu$ and $\rho_\mu^a$ are relevant.
With the above approximation, the equation of motion for the nucleon from the Lagrangian density (\ref{A1}) takes the form,
\begin{align} \label{A11}
    \Big(i{\g.\partial} - \Big(M_N - \tilde{g}_\s({\s}){\s} - \tilde{g}_\d({\d})\frac{{\tau}^3}{2}{{\d^3}}\Big) - \g^0\Big(g_\o \o_0 + g_\rho \frac{{\tau}^3}{2}\rho_{0}^3\Big) \Big)\psi &= 0. 
\end{align}
From eq. (\ref{A11}), the expression for the effective nucleon masses can be obtained as,
\begin{align}
\label{A3}
 M_{i = p,n}^*({\sigma},{\d^3}) = M_i - \tilde{g}_\s({\s}){\s} \mp \frac{1}{2}\tilde{g}_\d({\d}) {{\d}}^3.  
\end{align} 
The expectation values of the meson fields, under the MFA, can be derived by minimizing the energy density and can be expressed as follows:
\bea \label{A5}
\f &=& {\frac{1}{m_\f^2}\sum_{i=p,n} \left(-\frac{\partial M_i^*(\f)}{\partial \f}\right)\rho^{s}_i},\;\;{\rm{where,}}\;\f = \s,\d^3,\\
\label{A6}
\o_0 &=&\sum_{i=p,n} \frac{g_\o}{m_\o^2}\rho_{i} =\frac{g_\o}{m_\o^2}\rho_{B},\\ 
\label{A7}
\rho_0^3 &=& \sum_{i=p,n} \frac{g_\rho}{m_\rho^2}\frac{(\tau^3\rho)_{i}}{2} = -\frac{g_\rho}{m_\rho^2} \eta \rho_B,
\eea
with the scalar densities and the number densities of the nucleons given as,
\bea
\label{Arho}
{{\rho^s_i}} &=& \frac{2}{(2\pi)^3} \int d^3\vec{k} \Theta(k_{Fi} - |\vec{k}|) 
\frac{M_i^*(\s,\d^3)}{\sqrt{M_i^{*2}(\s,\d^3) + 
|\vec{k}|^2}}, \\ 
{{\rho_{i}}}&=& \frac{2}{(2\pi)^3} \int d^3\vec{k} \Theta(k_{Fi} - |\vec{k}|) = {{\frac{k_{Fi}^3}{3\pi^2}}}.
\eea
Here, $k_{Fi}$ denotes the nucleon ($i=p,n$) Fermi momentum and the factor of 2 represents the spin degeneracy factor. In eq. (\ref{A7}), $\eta$ is the isospin asymmetry parameter, defined as $\eta = (\rho_{n} - \rho_{p})/{(2 \rho_B)}$, where $\rho_B (= \sum_i \rho_i)$ represents the baryon density and $\rho_p$ and $\rho_n$ are the number densities of the proton and neutron, respectively.
The above expressions are at zero temperature, where the particle contributions reduce to the $\Theta$ function and the anti-particle contributions become zero \cite{menezes}. 
From eq. (\ref{A5}), it is observed that the scalar fields (${\f = \s,\d}$) exhibit self-consistent interactions within the nuclear matter. The scalar fields satisfy the self consistent relation in the following way,
\begin{eqnarray} \label{A12}
- \frac{\partial  M_i^*(\f) }{\partial \f} &=& \tilde{g}_\f(\f) = g_\f(\f=0)
C_i(\f),\\
\label{A13}
C_i(\f) &=& \frac{\sum_q n_{qi} S_{qi}(\f)}{\sum_q n_{qi} S_{qi}(\f=0)},
\end{eqnarray} 
where, $n_{qi}$ is the number of light valence (anti)quark constituents and $S_{qi}(\f)$ represents the scalar density of the light (anti)quarks ($q = u,\bar{u},d,\bar{d}$) of the respective hadron ($i$).
At the hadronic level, all the information about the constituent quarks is encrypted within this term. It might be noted that, for $C(\f)=1$, the QMC model turns into the Walecka model \cite{walecka}, where the nucleons are the fundamental degrees of freedom, interacting with the nucleons through mesons.
The meson-nucleon couplings $(g_\s,\,g_\d,\;g_\o,\;g_\rho)$ are defined based on 
quark-meson coupling strengths
$(g_\s^q,\;g_\d^q,\;g_\o^q\;g_\rho^q)$ as \cite{guichon1996}:
\begin{eqnarray} \label{A14}
g_\f (\f =0) &=& g_\f^q \sum_q n_{qi} S_{qi}(\f=0),\;
\label{A16}
g_\o = g_\o^q\sum_q n_{qi},\;g_\rho = g_\rho^q.
\label{gNs}
\end{eqnarray}
The equations of motion for the constituent light
(anti)quark fields inside the hadron bag ($i$) are represented as,
\bea
\label{B17}
&\Big[i\g.\partial - \Big(m_{u} - V_{\sigma} \mp \frac{1}{2}V_{\delta} \Big) \mp \g^0\Big(V_{\omega} + \frac{1}{2} V_{\rho}\Big)\Big] \begin{pmatrix}
    \psi_{ui}  \\
    \psi_{\bar{ui}} 
\end{pmatrix} = 0&,\\
\label{B18}
&\Big[i\g.\partial - \Big(m_{d} - V_{\sigma} \pm \frac{1}{2}V_{\delta} \Big) \mp \g^0\Big(V_{\omega} - \frac{1}{2} V_{\rho}\Big)\Big] \begin{pmatrix}
    \psi_{di}  \\
    \psi_{\bar{di}}  
\end{pmatrix} = 0&,
\eea
where, $V_\sigma = g_{\sigma}^q \sigma$, $V_{\delta} = g_{\delta}^q\delta^{3}$, $V_{\omega} = g_\omega^q \omega_0$ and $V_{\rho} = g_{\rho}^q\rho_{0}^3$ are the mean field potentials. The scalar and vector fields $\s$, $\d^3$, $\o_0$, and $\rho_{0}^3$ are obtained using the eqs. (\ref{A5})-(\ref{A7}) for given values of $\rho_B$ and $\eta$. Here, $m_u(=m_{\bar{u}})$, $m_d(=m_{\bar{d}})$ and $m_Q$ refer to the current quark masses and the expressions for the effective masses of the light quarks and antiquarks are obtained as,
\begin{eqnarray}
\label{B20}
m_{u}^\star &=& m_{u} - V_\sigma - \frac{1}{2} V_\delta,\\ \label{B21} m_{\bar{u}}^\star &=& m_{\bar{u}} - V_\sigma + \frac{1}{2} V_\delta,\\ \label{B22} m_{d}^\star &=& m_d - V_\sigma + \frac{1}{2} V_\delta, \\ \label{B23} m_{\bar{d}}^\star &=& m_{\bar{d}} - V_\sigma - \frac{1}{2} V_\delta.
\end{eqnarray}
The (anti)quark effective masses are modified by the mean scalar potentials, where $\d^3$ breaks the isospin symmetry of quark and antiquark doublets and causes $u$ and $d$ quarks as well as the antiquarks to behave differently in the asymmetric nuclear medium (ANM). As a consequence, the light (anti)quark doublets exhibit a mass splitting within the isodoublets in ANM. However, the heavy quark ($Q=s,c,b,\bar{s},\bar{c},\bar{b}$ ) masses are assumed to be the same in the nuclear medium as in the free space ($m^\star_{Q} = m_{Q}$), since, in the QMC model, the mean fields are coupled only to the light quarks and light antiquarks.
In eqs. (\ref{B17}) and (\ref{B18}), $\psi_{fi}$ represents the static ground state wave function for the (anti)quark fields of flavor $f(=q,Q)$ in the bag, and, is given by
\begin{align}
\label{B25}
\psi_{fi} ({r}) = N_{fi} e^{- \frac{i \epsilon^\star_{fi} t}{ R^\star_i}}\begin{pmatrix}
    j_0(\frac{x^\star_{fi} r}{R_i^\star}) \\
    i\beta^\star_{fi}\vec{\sigma}.\hat{r} j_1(\frac{x^\star_{fi} r}{R^\star_i})
\end{pmatrix} \frac{\chi_{fi}}{\sqrt{4\pi}}, \; |\vec{r}| \leq \rm{Bag\; radius}\;(R_i),
\end{align}
where $N_{fi}$ is the normalization factor \cite{santos}, $j_0$ and $j_1$ are the spherical Bessel functions, $R_i^\star$ is the in-medium bag radius and  $\chi_{fi}$ refers to the quark spinors. The energy eigenvalues (in units of $1/R_i^\star$) of the quarks and antiquarks, which are denoted by $\epsilon_{fi}^\star$ in eq. (\ref{B25}), are given as,
\begin{eqnarray}
\label{B26}
\left( \begin{array}{c} \e^\star_{ui} \\ \e^\star_{\bar{ui}} \end{array} \right)
&=& \Omega_{qi}^\star \pm R^\star_i \left( 
V_\omega + \frac{1}{2} V_\rho \right),\\ \label{B27}
\left( \begin{array}{c} \e^\star_{di} \\ \e^\star_{\bar{di}} \end{array} \right)
&=& \Omega_{qi}^\star \pm R^\star_i \left( 
V_\omega - \frac{1}{2} V_\rho \right),\\ \label{B28}
\e^\star_{Qi} &=& \e_{Qi}= \Omega^\star_{Qi}.
\end{eqnarray}
The other parameters in eq. (\ref{B25}) are, 
\begin{align}
\label{B29}
\beta^\star_{fi} = \sqrt{\frac{\Omega_{fi}^\star - R^\star_i m_{fi}^\star}{\Omega_{fi}^\star + R^\star_i m_{fi}^\star}}\;,\;\;\; {\rm{where,}}\;\;\; 
\Omega_{fi}^\star = \sqrt{x_{fi}^{\star 2} + (R^\star_i m_{fi}^\star)^2}.
\end{align}
The parameters $x_{fi}$'s are the bag eigen frequencies in units of $1/R_i^\star$ \cite{MIT}, determined by the linear boundary condition at the bag surface ($|\vec{r}| = R_i$) \cite{thomas1982},
\begin{align}
    \label{B31}
    i\g^{\mu} n_{\mu} \psi_{fi} = \psi_{fi}\; \implies\;j_0(x^\star_{fi}) = \beta^\star_{fi} j_1 (x^\star_{fi}).
\end{align}
In the QMC model, the effective mass of the hadrons in the nuclear medium at rest is equal to the energy of the static bag consisting of ground state quarks and antiquarks, which can be obtained as,
\begin{align}
    \label{B32}
    M_i^\star(\s,\d^3) =E_i^{bag}(\s,\d^3)= \frac{\sum_f{n_{fi}\Omega_{fi}^\star}-Z_i}{R^\star_i} + \frac{4}{3} \pi R_i^{\star 3} B,
\end{align}
where the parameter $Z_i$ provides the correction regarding the centre of mass motion and the gluon fluctuations and $B$ is the bag pressure which provides an inward pressure to balance the outward pressure exerted by the motion of the quarks and antiquarks inside the bag. As a consequence, there is an equilibrium where the hadron mass
is minimized and stabilized, as follows :
\begin{align}
    \label{B33}
    \frac{d M_i^\star}{d R_i}{\Big{|}}_{R_i^\star} = 0.
\end{align}
Using the above condition the in-medium bag radius, $R_i^\star$, can be determined. It is important to remember that the bag radius is different from the actual hadron radius, which can be computed from the quark wave functions. 

The scalar density $S(\f)$, $\phi =\s,\d^3$, given in eq. (\ref{A13}) is expressed in terms of the scalar densities of the light (anti)quarks in the bag, which are given as \cite{niu},
\bea \label{B34}
S_{qi}(\s)&=& \mathcal{I}^s_i ,\;\;\;\;S_{qi}(\d^3) = \frac{\tau_q^3}{2}\mathcal{I}^s_i,\\
\mathcal{I}^s_i &=& \int d\vec{r}\bar{\psi}_{qi} \psi_{qi} 
\nonumber\\ &=&  \frac{\Omega_{qi}/2 + m_{qi}^\star R^\star_i(\Omega_{qi}-1)}{\Omega_{qi}(\Omega_{qi}-1) + m_{qi}^* R^\star_i /2},
\eea
The above term plays an important role in this model, where it introduces the non-linearity and the self-consistency of the scalar meson fields and thus constitutes the new saturation mechanism of nuclear matter based on the quark degrees of freedom.
Eq. (\ref{A12}), for the scalar fields and the nucleons ($i =p,\;n$) can be explicitly written as,
\bea
\label{DM1}
- \frac{\partial  M_p^*(\f) }{\partial \s} & =& g_\s^q( 2 S_{up}+S_{dp}) = g_\s^q S_{\s}^p,\\ \label{DM2}
- \frac{\partial  M_n^*(\f) }{\partial \s} & =& g_\s^q( S_{un}+2S_{dn}) = g_\s^q S_{\s}^n,\\ \label{DM3}
- \frac{\partial  M_p^*(\f) }{\partial \d^3} & =& \frac{g_\d^q}{2}( 2 S_{up}-S_{dp}) = \frac{g_\d^q}{2}S_{\d}^p,\\ \label{DM4}
- \frac{\partial  M_n^*(\f) }{\partial \d^3} & =& \frac{g_\d^q}{2}( S_{un}-2 S_{dn}) = \frac{g_\d^q}{2}S_{\d}^n,
\eea
and hence the mean scalar meson fields ($\f$) of eq. (\ref{A5}) turn out to be,
\bea
\label{sigma}
\s &=&\frac{1}{m_\s^2}\times \Big[\rho^s_p \left(-\frac{\partial M_p^*(\f)}{\partial \s}\right) + \rho^s_n \left(-\frac{\partial M_n^*(\f)}{\partial \s}\right)\Big]\nonumber\\
&=&\frac{g_\s^q}{m_\s^2}\times \Big[\rho^s_p (S_{\s}^p) + \rho^s_n ( S_{\s}^n)\Big],
\\ \label{delta}
\d^3 &=&\frac{1}{m_\d^2}\times \Big[\rho^s_p \left(-\frac{\partial M_p^*(\f)}{\partial \d^3}\right) + \rho^s_n \left(-\frac{\partial M_n^*(\f)}{\partial \d^3}\right)\Big]\nonumber\\
&=&\frac{g_\d^q}{2 m_\d^2}\times \Big[\rho^s_p (S_{\d}^p) + \rho^s_n ( S_{\d}^n)\Big].
\eea
\section{OPEN STRANGE AND OPEN HEAVY flavor MESONS IN NUCLEAR MEDIUM}
\label{3}
In this section, we present the impacts of medium-modified quark properties on the open strange and open heavy flavor mesons in the asymmetric nuclear medium. We study the in-medium properties, e.g., masses and excitation energies, of $K(\bar{K})$, $D(\bar{D})$, and $B(\bar{B})$ mesons within the QMC$\d$ model. For investigating the properties of these mesons, we can express the interactions of these mesons within the nuclear medium using the QMC model, where the light quark(antiquark) constituents of these pseudoscalar mesons ($P\equiv K,\bar{K}, D,\bar{D}, B,\bar{B}$) are subjected to the scalar ($V_\s, V_\d$) as well as vector ($V_\o, V_\rho$) potentials generated within the medium. The corresponding interaction Lagrangian \cite{tsushima1,schaffner} is given as
\begin{align} \label{37}
    \mathcal{L}_P = (D_\mu {{P}})^\dagger D^\mu P - M_P^{\star 2} {\bar{P}}P - g^q_{P\s} M^\star_P {\bar{P}} P{\s} - g^q_{P\d}M^\star_P {\bar{P}}\frac{{\tau}^a}{2} P {{\d}}^a,
\end{align}
where the vector meson-pseudoscalar meson couplings are introduced through the covariant derivative, $D_\mu = \partial_\mu + i g^q_{P\o} {\o}_\mu + i g^q_{P\rho}\frac{\tau^a}{2}{{\rho}}_{\mu}^a$, and $g^q_{P\s}$, $g^q_{P\d}$, $g^q_{P\o}$ and $g^q_{P\rho}$ represent the couplings between the meson fields and the pseudoscalar meson ($P$) at the quark level. The effective mass of the pseudoscalar ($P$) meson, $M_P^*$, is obtained from eq. (\ref{B32}), where the subscript $i$ for hadron refers to the pseudoscalar meson $P$.
To study the excitation energies, we consider the meson ($P$) as a static spherical MIT bag, whose light valence quark contents are allowed to interact with the scalar ($\s,\;{\d}$) as well as vector ($\o,\;{\rho}$) meson fields.
These coupling constants are related to the quark-meson couplings in the following way \cite{saito2}
\begin{eqnarray} \label{38}
    g_{P\s}^q &=& \frac{n_q}{3}g^q_\s,\\ \label{39}
    g^q_{P\d} &=& \frac{n_q}{3} g^q_\d,\\ \label{40}
    g_{P\o}^q &=& 1.4^2g_\o^q,\\ \label{41}
    g_{P\rho}^q &=& g_\rho^q,
\end{eqnarray}
where $n_q$ denotes the count of valence light quarks (antiquarks) in the respective hadron. The vector coupling constant, $g_\o^q$ is rescaled by a factor of $(1.4)^2$, to be consistent with the empirically obtained $K^+ N$ scattering data \cite{tsushima1}
and it is also chosen to be the same way for open charm and open bottom mesons \cite{tsushimaC}. Considering the above interactions, the equation of motion for $P$ meson has the general form,
\begin{align}
    \label{42}
    \Big(\partial_\mu\partial^\mu + M_P^{\star 2} + M_P^\star(g^q_{P\s}\s + g^q_{P\d}\frac{{\tau}^a}{2}\d^{a}) + 2(g^q_{P\o} \o_{\mu} + g^q_{P\rho} \frac{{\tau}^a}{2}\rho_{\mu}^a)i \partial^\mu \nonumber\\- (g^q_{P\o} \omega_\mu + g^q_{P\rho} \frac{{\tau}^a}{2}\rho_{\mu}^a )^2\Big) P = 0.
\end{align}
Using a plane wave expansion of the $P$ meson and
taking the Fourier transformation of the equation of motion, we obtain the dispersion relations for the pseudoscalar mesons in the mean field approximation.
The in-medium excitation energies \cite{tsushima1} of the pseudoscalar mesons ($\o_P$) are obtained from the solution of the dispersion relations for $\vec{k} = 0$, which are given as,
\begin{eqnarray}
    \label{44}
    \left( \begin{array}{c} \o_{K^+} \\ \o_{K^-} \end{array} \right)
&=& \left( \begin{array}{c} \sqrt{M^{\star 2}_{K^+} + M^\star_{K^+}\tilde{V}_\s^+ } \\ \sqrt{M^{\star 2}_{K^-} + M^\star_{K^-}\tilde{V}_\s^-}\end{array} \right) \pm \left( 
\tilde{V}_\o + \frac{1}{2} V_\rho \right) \\ \label{45}
\left( \begin{array}{c} \o_{K^0} \\ \o_{\bar{K}^0} \end{array} \right)
&=& \left( \begin{array}{c} \sqrt{M^{\star 2}_{K^0} + M^\star_{K^0}\tilde{V}_\s^- } \\ \sqrt{M^{\star 2}_{\bar{K}^0} + M^\star_{\bar{K}^0}\tilde{V}_\s^+} \end{array} \right) \pm \left( 
\tilde{V}_\o - \frac{1}{2} V_\rho \right)  \\ \label{46}
    \left( \begin{array}{c} \o_{D^+} \\ \o_{D^-} \end{array} \right)
&=& \left( \begin{array}{c} \sqrt{M^{\star 2}_{D^+} + M^\star_{D^+}\tilde{V}_\s^+} \\ \sqrt{M^{\star 2}_{D^-} + M^\star_{D^-}\tilde{V}_\s^- } \end{array} \right) \mp \left( 
\tilde{V}_\o - \frac{1}{2} V_\rho \right) \\ \label{47}
\left( \begin{array}{c} \o_{D^0} \\ \o_{\bar{D}^0} \end{array} \right)
&=& \left( \begin{array}{c} \sqrt{M^{\star 2}_{D^0} + M^\star_{D^0}\tilde{V}_\s^-}\\ \sqrt{M^{\star 2}_{\bar{D}^0} + M^\star_{\bar{D}^0}\tilde{V}_\s^+ } \end{array} \right) \mp \left( 
\tilde{V}_\o + \frac{1}{2} V_\rho \right) \\ \label{48}
    \left( \begin{array}{c} \o_{B^+} \\ \o_{B^-} \end{array} \right)
&=& \left( \begin{array}{c} \sqrt{M^{\star 2}_{B^+} + M^\star_{B^+}\tilde{V}_\s^+ } \\ \sqrt{M^{\star 2}_{B^-} + M^\star_{B^-}\tilde{V}_\s^-}\end{array} \right) \pm \left( 
\tilde{V}_\o + \frac{1}{2} V_\rho \right) \\ \label{49}
\left( \begin{array}{c} \o_{B^0} \\ \o_{\bar{B}^0} \end{array} \right)
&=& \left( \begin{array}{c} \sqrt{M^{\star 2}_{B^0} + M^\star_{B^0}\tilde{V}_\s^- } \\ \sqrt{M^{\star 2}_{\bar{B}^0} + M^\star_{\bar{B}^0}\tilde{V}_\s^+} \end{array} \right) \pm \left( 
\tilde{V}_\o - \frac{1}{2} V_\rho \right) 
\end{eqnarray}
where
$\tilde{V}_\s^\pm = (g^q_{P\s}\s \pm \frac{1}{2}g^q_{P\d} \d^3)\;
{\rm {and}}\;\tilde{V}_\o = 1.4^2 V_\o$.

\section{Results and Discussions}
\label{4}
In the present work, we study medium modifications of masses and excitation energies of open strange ($ K^+,\;K^-,\;K^0,\;\bar{K}^0$), open charm ($D^+,\;D^-,\;D^0,\;\bar{D}^0$), and open bottom ($ B^+,\;B^-,\;B^0,\;\bar{B}^0$) mesons in both the symmetric nuclear matter (SNM) as well as in the asymmetric nuclear matter (ANM) using the QMC$\d$ model. It is assumed that the in-medium masses of the pseudoscalar mesons are generated through the interactions of the constituent light quarks (antiquarks) with the scalar meson fields ($\sigma$ and $\delta$) through the scalar density of (anti)quarks, given by eq. (\ref{B34}). On the other hand, the excitation energies are modified by both the scalar and vector meson fields using eqs. (\ref{44})-(\ref{49}). The parameters of the model as well as the in-medium properties of the pseudoscalar mesons are presented in the following subsections.
\subsection{Parameters of QMC$\d$ model} \label{4A}
We first state the parameters of the model we have chosen in the present work before discussing the results. The bag parameters $B$ and $Z$ are fitted from the mass and the given bag radius of the proton in the free space, using the eqs. (\ref{B32}) and (\ref{B33}). 
It might be noted here that we have considered in the present study, the small isospin breaking effect between proton and neutron due to the electromagnetic corrections \cite{pdg}. Using slightly different masses(in MeV) of $u$ and $d$ quarks, $m_u = 2.16$, $ m_d = 4.67$, for the heavier quarks  $ m_s = 93.4$, $m_c = 1270$, and $m_b = 4180$ \cite{pdg} and the bag constant, $B = (211.2378\; \rm{MeV})^4$ the values of $R$ and $Z$ are calculated. These are shown in table - \ref{t1}.

\begin{table}[th]
\centering
\begin{tblr}{colspec={ccccccc}}
\hline
& $M$(MeV) (I) &$R$ (fm)& $Z$ & $M^\star$(MeV) & $R^\star$ (fm)\\
\hline
$p$ & 938.272&  0.6 (I) &  4.0015 &733  & 0.59\\
$n$&  939.565 &  0.6003 & 4.0012 &  719 & 0.58\\
\hline
$K^0\; ({\bar{K}^0})$ &  497.611&  0.4824 & 3.2860 & 419.7 (434.3) & 0.4771 (0.4790)\\
$K^+\; (K^-$) &  493.677 &  0.4811 &  3.2924 & 430.5 (415.9) & 0.4775 (0.4756)\\
\hline
$D^0\; ({\bar{D}^0})$ &  1864.84&  0.5798 &  1.8423 & 1788.3 (1802.4) & 0.5758 (0.5772)\\
$D^+\; ( D^-)$ &  1869.66 &  0.5808 &  1.8319 & 1807.1 (1792.9) & 0.5782 (0.5769)\\
\hline
$B^0\; ({\bar{B}^0})$ &  5279.66&  0.6508 &  -0.2609 & 5204.1 (5217.9) & 0.6470 (0.6482)\\
$B^+\; ( B^-)$ &  5279.34 &  0.6507 &  -0.2638 & 5217.7 (5204.0)& 0.6483 (0.6469)\\
\hline
\end{tblr}
\caption{ \raggedright{Representative inputs (I), parameters that are used in the present study and some relevant properties calculated at nuclear saturation density, $\rho_0$ (denoted by asterisk symbol) in the asymmetric nuclear matter. The chosen values of current quark masses (in MeV) are} $m_u = 2.16$, $ m_d = 4.67$, $ m_s = 93.4$, $m_c = 1270$, and $m_b = 4180$ \cite{pdg} and the calculated bag constant, $B = (211.2378\; \rm{MeV})^4$}
\label{t1}
\end{table}
\begin{table}[th]
\centering
\begin{tabular}{ c   c   c   c }
\hline
 $g_\s^q$ \qquad & \qquad $g_\d^q$ \qquad & \qquad $g_\o^q$ \qquad & \qquad $g_\rho^q$\\
\hline
5.98 \qquad &\qquad  12.60 \qquad &\qquad  2.98 \qquad &\qquad  12.59\\
\hline
\end{tabular}
\caption{ \raggedright{Values of quark-meson-field coupling constants \cite{santos}}}
\label{t2}
\end{table}
Next, with the meson masses taken as $m_\s = 550 \;\rm{MeV}$, $m_\o = 783 \;\rm{MeV}$, $m_\rho = 770 \;\rm{MeV}$ and $m_\d = 983 \;\rm{MeV}$, we fix the nucleon meson coupling constants $g_\s,\;g_\o,\;g_\rho$, {\rm{and}}\; $g_\d$
using the bulk properties of nuclear matter such as binding energy per nucleon ($-15.7\;{\rm{MeV}}$), incompressibility ($200-300\;{\rm{MeV}}$), symmetry energy ($32.5\;{\rm{MeV}}$), and slope of the symmetry energy ($88\pm25 \;{\rm{MeV}}$), at saturation density ($0.15\; {\rm{fm^{-3}}}$) \cite{santos}.
From these, the quark meson coupling constants are obtained using relation (\ref{A14}), as shown in table - \ref{t2}. As has been stated earlier, for investigating the properties of open strange, charm, and bottom mesons with the vector meson fields we take $g_{P\o}^q = 1.4^2 \times g_\o^q$.
\subsection{Meson (scalar and vector) fields and quark masses in nuclear medium} \label{4B}
\begin{figure}[th]
\centerline{\includegraphics[width=\textwidth]{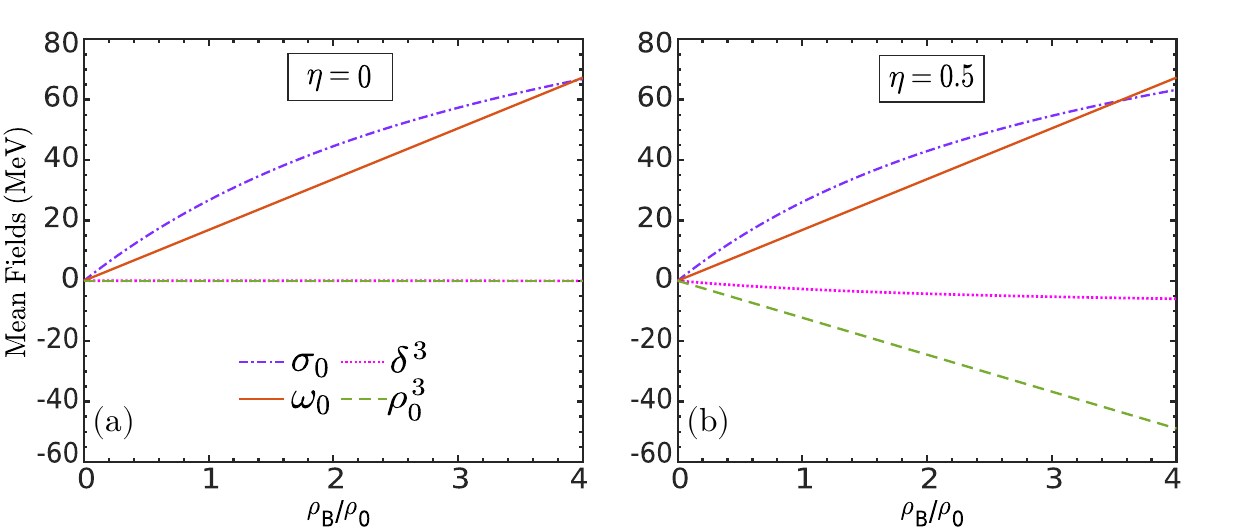}}
\vspace*{8pt}
\caption{\raggedright{Behaviors of the mean meson fields as a function of baryon density $\rho_B$, in the units of saturation density ($\rho_0 = 0.15\; {\rm{fm^{-3}}}$). The results are shown for both the (a) symmetric ($\eta = 0$) and (b) asymmetric ($\eta = 0.5$) nuclear matter.}}\label{fig1}
\end{figure}

We first examine the behavior of the mean scalar and vector meson fields, which interact with the light (anti)quarks in the dense symmetric nuclear matter (SNM) and asymmetric nuclear matter (ANM) within the QMC$\d$ model. The meson fields are determined by solving the coupled nonlinear eqs. (\ref{A5})-(\ref{A7}) using the mean field approximation. Fig. \ref{fig1} illustrates the density dependence of the mean meson fields, where $\eta = 0$ and $\eta = 0.5$ represent the SNM and the highest degree of asymmetry which corresponds to pure neutron matter, respectively.
\begin{figure}[th]
\centerline{\includegraphics[width=10cm,height=14cm]{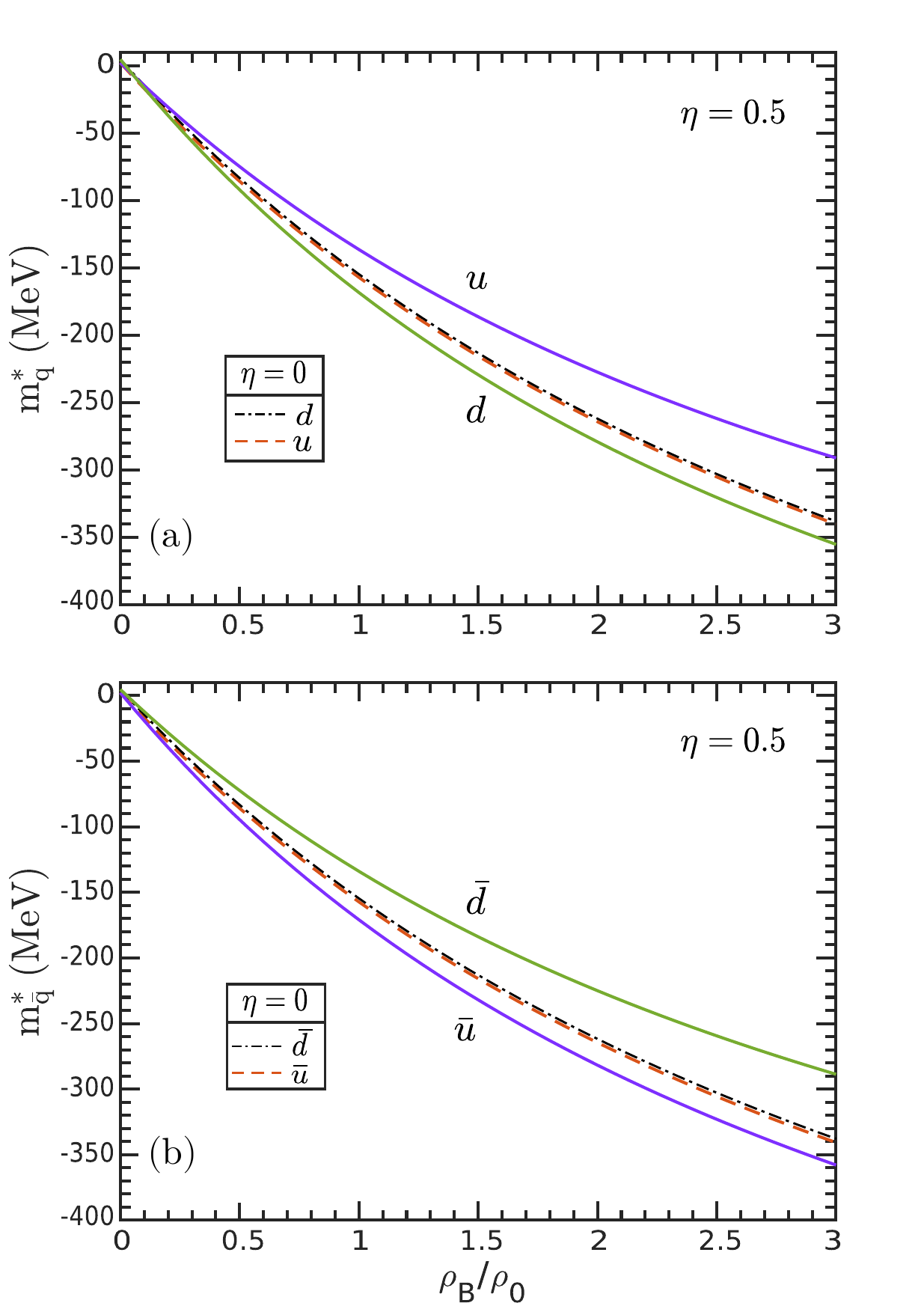}}
\vspace*{8pt}
\caption{\raggedright{Masses of light (a) quarks and (b) antiquarks with respect to baryon density, in the units of saturation density ($\rho_0 = 0.15\; {\rm{fm^{-3}}}$). Broken lines represent the masses in symmetric nuclear matter. Dashed lines : ($u$, $\bar{u}$), dash-dotted lines : ($d$, $\bar{d}$), solid lines : $\eta = 0.5$.}}\label{figq}
\end{figure}
As can be seen from the equations (30)-(35),
the expectation value of $\s$ ($\d$) meson field is obtained from the sum (difference) of positive quantities, i.e.,  $(2S_{up} + (-) S_{dp})$  for proton
and $(S_{un} + \small(-) 2S_{dn})$ for neutron.
As a consequence of this, the $\d$ meson attains much smaller expectation values as compared to the $\s$ meson in the asymmetric dense nuclear matter as depicted in fig. \ref{fig1}.
In the asymmetric nuclear matter, due to the presence of the $\delta$ field, the expectation value of the $\sigma$ field decreases in the high-density regions as compared to SNM. This is because, these two fields are coupled to each other (see eq. (\ref{A5})), and the obtained expectation values of these fields are dependent on the nucleon scalar densities ($\rho_i^s$), which are different for different values of isospin asymmetry parameter $\eta$.
On the other hand, the expectation value of the $\rho$ meson depends on ($\rho_p -\rho_n$). Therefore, the matter with a neutron excess induces a negative isovector density, which gives rise to the negative in-medium expectation value of the mean vector isovector meson field at finite densities in the ANM. 
Within the QMC$\d$ model, the interaction of light quarks and antiquarks with these mean meson fields results in modifications of the quark and antiquark properties in the nuclear medium. The scalar fields influence their masses (see eqs. (\ref{B20})-(\ref{B23})), while the vector fields, together with the scalar fields, alter the energies of the (anti)quarks. The graphical representation of the in-medium behaviors of the light quarks' ($u,\;d$) and light antiquarks' ($\bar{u},\;\bar{d}$) masses on the baryon density are shown in fig. \ref{figq}. 
In SNM, due to the attractive potential exerted by the scalar-isoscalar field, there is observed to be a reduction in the masses of quarks and antiquarks in the nuclear medium. 
In ANM, the inclusion of the $\d$ meson field leads to the breaking of isospin degeneracy of the masses of the light quark and antiquark doublets and causes mass splitting between ($u,\;d$) as well as ($\bar{d},\;\bar{u}$). At saturation density ($\rho_0$) it is visible from fig. \ref{figq} that there is almost 32 MeV difference between $u$ and $d$ quark masses and around 37 MeV difference between $\bar{u}$ and $\bar{d}$ masses, which increases with the increase in density.
Such considerable mass splittings can be understood from eqs. (\ref{B20})-(\ref{B23}), where
the $\d$ meson induces opposite interactions between the (anti)quarks within the isospin doublets and hence leads to a repulsive contribution to the $u(\bar{d})$ mass and attractive contribution for $d(\bar{u})$ mass. As a result, the mass drop of the $u$ quark becomes lesser than that of the $d$ quark in ANM, and the mass drop of $\bar{u}$ quark (in SNM which is identical to $u$ quark) becomes higher than the mass shift of $\bar{d}$ quark (which is identical to $d$ quark in SNM), as can be seen from fig. \ref{figq} and which lead to significant mass splittings between the light (anti)quarks. 
Furthermore, it is important to note that as we have considered small current masses for the (anti)quarks, the obtained values of effective (anti)quark masses in the nuclear medium are negative, which indicates the attractive nature of scalar potentials, as stated in eqs. (\ref{B17}) and (\ref{B18}) in section {\ref{2}}. Therefore, the naive interpretation of the physical mass for a particle, which is positive, should not be applied for $m_q^*$, which in any case is not observable \cite{tsushima19}. 
\subsection{Masses of open strange and open heavy flavor mesons in nuclear medium} \label{4C}
\begin{figure}[th]
\centerline{\includegraphics[width=\textwidth]{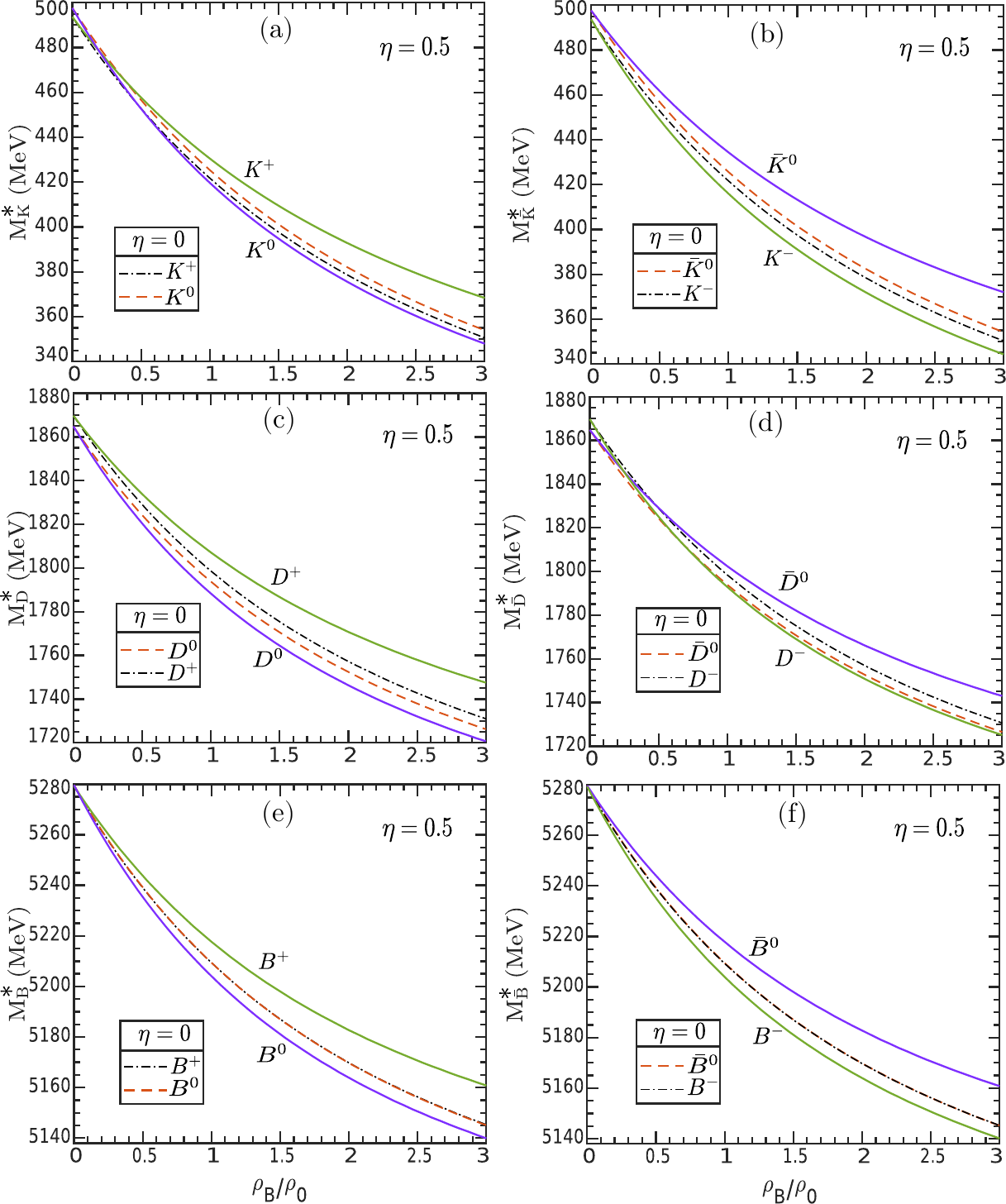}}
\vspace*{8pt}
    \caption{\raggedright{The effective masses of (a) $K$ (b) $\bar{K}$ (c) $D$ (d) $\bar{D}$ (e) $B$ and (f) $\bar{B}$  mesons as a function of baryon density $\rho_B$ (in the units of $\rho_0$). Solid lines: $\eta = 0.5$. Broken lines: $\eta = 0$. }}
    \label{fig2}
\end{figure}

In this sub-section, we present the results for in-medium masses of $K(\bar{K})$, $D(\bar{D})$ and $B(\bar{B})$ mesons in symmetric as well as asymmetric nuclear matter. The properties of these mesons in nuclear matter undergo medium modifications due to the interactions of their constituent light (anti)quarks with the mean meson fields within the QMC$\d$ model.
Therefore the scalar fields attaining different values from their vacuum values at finite densities leads to mass modifications of these mesons, which are shown in fig. \ref{fig2}.
We observe that the in-medium masses of the respective mesons decrease nonlinearly with the increase of baryon density, $\rho_B$. 
In symmetric nuclear matter (SNM), the $K(\bar{K})$, $D(\bar{D})$ and $B(\bar{B})$ mesons follow a similar decreasing pattern, accompanied by a small isospin symmetry breaking contribution due to their different charges (see section \ref{4A}). However, one can see from fig. \ref{fig2} that, the effective masses of the members in the isospin doublets
have a splitting in the asymmetric nuclear matter (ANM). This is due to the incorporation of the scalar-isovector $\d$ meson field in the model. The $\d$ meson treats the constituent light quarks i.e., up and down quarks(antiquarks) differently as might be seen from eqs. (\ref{B20})-(\ref{B23}) and the mass splitting in the asymmetric nuclear matter between them is illustrated in fig. \ref{figq}.
In SNM, at $\rho_B =\rho_0$, both the $K^0(\bar{K}^0)$ and $K^+(K^-)$ meson masses are observed to undergo a reduction of approximately $14\%$ from their vacuum masses. In asymmetric nuclear matter, there is observed to be splitting between the masses of the isodoublets. In the context of the $K$ meson doublet, the $K^0(K^+)$ meson contains a light $d(u)$ quark, which has a drop (rise) in the ANM (see discussion in \ref{4B}). This leads to a splitting between the $K^0$ and $K^+$ mesons within the kaon isodoublet. Further, there are also contributions to the masses of the mesons wihtin an isospin doublet due to the different bag parameters of the charged and neutral mesons (see table - \ref{t1}). Therefore in ANM, there is a drop(rise) of $K^0(K^+)$ by about $1.4\%(2\%)$ and rise(drop) of $\bar{K}^0(K^-)$ of about $2\%(1.4\%)$ as compared to the value in symmetric nuclear matter at $\rho_B=\rho_0$. 
Figs. \ref{fig2}(a) and \ref{fig2}(b) distinctly demonstrate that the masses of $K^0(K^-)$ and $K^+(\bar{K}^0)$ shows a contrasting behavior in ANM, with $K^0(K^-)$ experiencing a decrease in mass and $K^+(\bar{K}^0)$ encountering an increase in mass relative to their values in SNM.
Figs. \ref{fig2}(c) and \ref{fig2}(d) display the in-medium behavior of the $D$ and $\bar{D}$ meson masses. In SNM, at $\rho_B = \rho_0$, both the charged and neutral $D$ meson masses exhibit a decrease of approximately $4\%$ from their vacuum masses.
For asymmetric nuclear matter ($\eta=0.5$), at $\rho_B=\rho_0$, the $D^0$ and $D^-$ meson masses undergo a decrease of around $0.3\%$, while the $D^+$ and $\bar{D}^0$ meson masses experience an increase of roughly $0.5\%$ as compared to their respective masses in SNM. 
As can be observed from figs. \ref{fig2}(e) and \ref{fig2}(f), due to the small vacuum mass differences between the charged and the neutral open bottom mesons, in SNM, the mass difference between masses of the members of the isodoublets are negligible.
In SNM, the masses of open bottom mesons decrease by around $1.3\%$ at nuclear saturation density. Similar to other pseudoscalar mesons, the $B$ and $\bar{B}$ mesons also experience mass splittings in ANM. 
The mass of the $B^0$ ($B^-$) meson experiences a reduction of approximately $0.1\%$, while the $B^+$ ($\bar{B}^0$) meson undergoes an increment of roughly $0.16\%$ at $\rho_B=\rho_0$, in ANM ($\eta = 0.5$) as compared to their masses in SNM. The in-medium masses of the studied mesons at the saturation density ($\rho_0$) are presented in the table - \ref{t1}.
The mass shift of $B^+$ and $B^0$ mesons are observed to be more, with increasing density, whereas the mass shifts for $B^-$ and $\bar{B}^0$ remain almost constant as compared to the values in symmetric nuclear matter. The percentage of reduction of masses in the medium is observed to be larger for the open strange mesons as compared to the open heavy flavor mesons, which is attributed to the increasing order of their masses in free space \cite{krein1}.

Based on the above analysis, it can be observed that the medium modifications for $K^+$, $\bar{K}^0$, $D^+$, $\bar{D}^0$, $B^+$ and $\bar{B}^0$ meson masses are more pronounced and repulsive as compared to their isospin counterparts. This disparity arises due to the individual nature of mean field potentials experienced by the mesons in the medium. Specifically, $K^+(u\bar{s})$, $\bar{K}^0(s\bar{d})$, $D^+(c\bar{d})$, $\bar{D}^0(u\bar{c})$, $B^+(u\bar{b})$ and $\bar{B}^0(b\bar{d})$ experience an attractive scalar isoscalar potential and a repulsive scalar isovector potential, whereas both the mean scalar field potentials are attractive for their isospin counterparts $K^0(d\bar{s})$, ${K}^-(s\bar{u})$, $D^0(c\bar{u})$, ${D}^-(d\bar{c})$, $B^0(d\bar{b})$ and ${B}^-(b\bar{u})$ mesons in the asymmetric nuclear matter.
The isospin effects are observed to be more pronounced at higher baryon densities.
Such strong isospin dependence on the in-medium masses of the pseudoscalar mesons should influence their production and propagation in the isospin asymmetric nuclear matter.
\begin{figure}[th]
\centerline{\includegraphics[width=\textwidth]{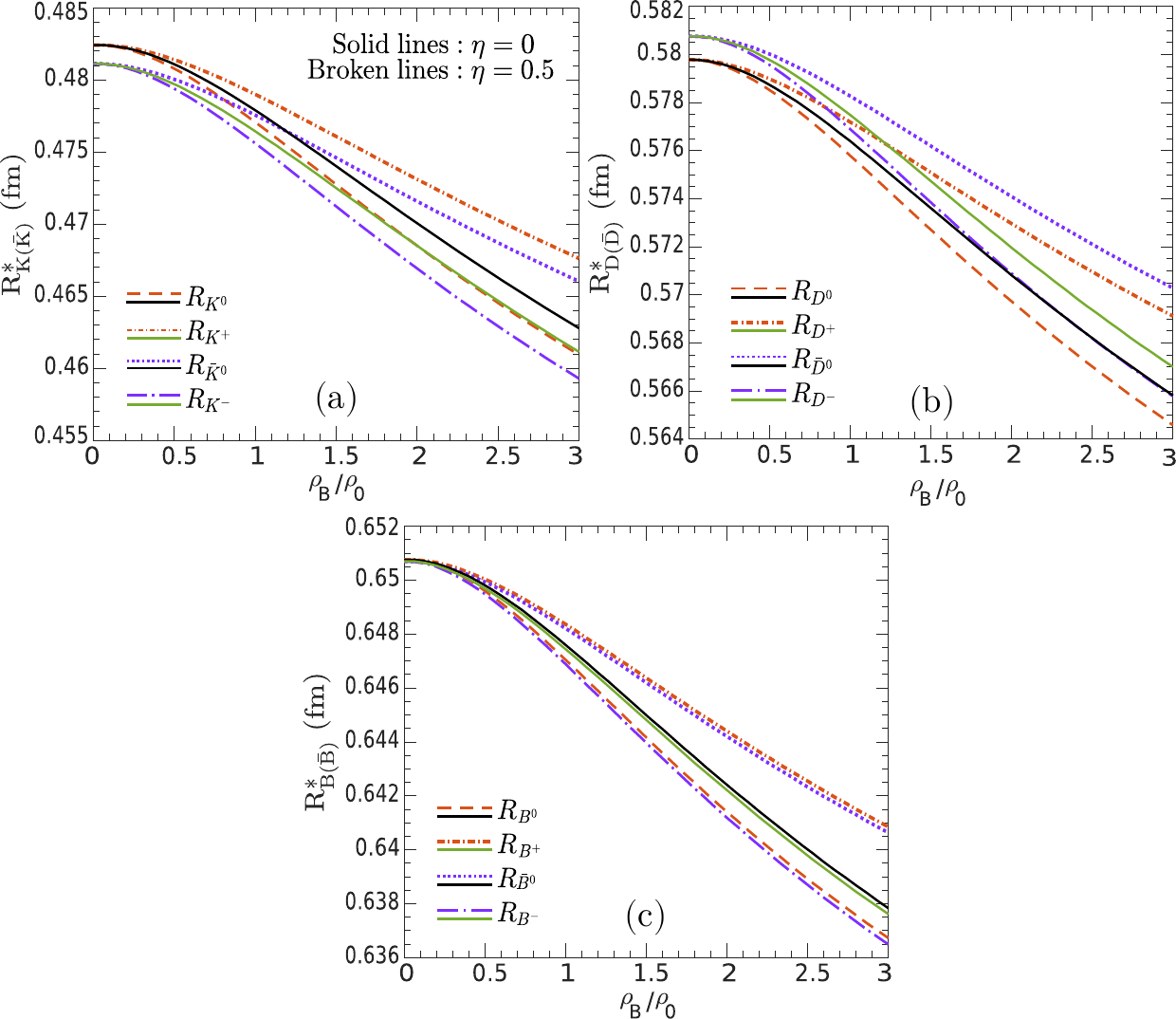}}
\vspace*{8pt}
    \caption{\raggedright{Variation of bag radii(in fm) for (a)$K(\bar{K})$ (b)$D(\bar{D})$ and (c)$B(\bar{B})$ mesons in the SNM (solid lines) as well as in ANM (broken lines), as a function of baryon density $\rho_B$ (in the units of $\rho_0$) in QMC$\d$ model.}}
    \label{figR}
\end{figure}

In the present study, we have also taken into account the medium modifications of the bag radii for the nucleons \cite{santos} and pseudoscalar mesons (see table - \ref{t1}). The behavior of the in-medium proton and neutron radii are already been discussed in the reference \cite{santos}. The temperature dependence of the bag radius of the nucleons has also been studied in the literature \cite{Mishra:1997kk}. Within the present model, the hadron bags in the symmetric as well as asymmetric nuclear matter are observed to shrink with the increase in density, as can be seen from fig. \ref{figR}.
In ANM, there occurs splitting of radii between the isospin doublets of the corresponding mesons, where the shift in radii of $K^0({K}^-)$, $D^0(D^-)$ and $B^0({B}^-)$ become greater than their respective isodoublet counterparts $K^+(\bar{K}^0)$, $D^+(\bar{D}^0)$ and $B^+(\bar{B}^0)$ due to the opposite effects arising from scalar isovector meson field. 
\subsection{Excitation energies of open strange and open heavy flavor mesons} \label{4D}
\begin{figure}[th]
\centerline{\includegraphics[width=\textwidth]{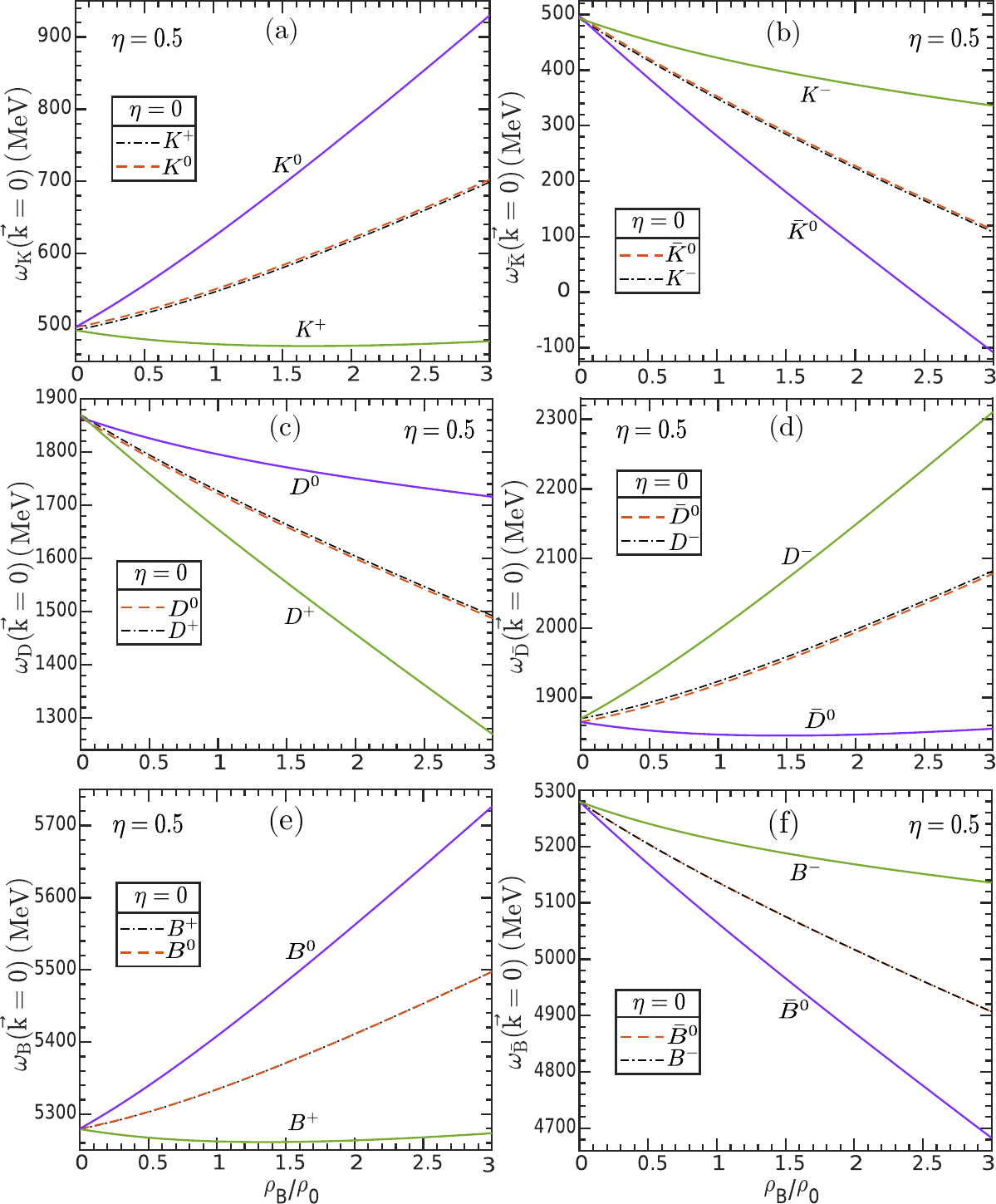}}
\vspace*{8pt}
    \caption{\raggedright{The excitation energies ($\o$) of (a) $K$ (b) $\bar{K}$ (c) $D$ (d) $\bar{D}$ (e) $B$ and (f) $\bar{B}$ mesons at $\vec{k} =0$ (in MeV) vs. baryon density $\rho_B$ (in the units of $\rho_0$), where the solid lines correspond to $\eta = 0.5$ and the broken lines correspond to symmetric nuclear matter ($\eta=0$). }}
    \label{fig3}
\end{figure}

We next analyze the behavior of the excitation energies ($\o_P$) of the pseudoscalar mesons ($P$) for $\vec{k}=0$ in the nuclear medium. (
In isospin asymmetric nuclear matter, the interactions between the mean scalar fields ($\s,\d$) as well as the mean vector ($\o,\rho$) fields with the (anti)quarks in the medium lead to the modifications of the excitation energies of these hadrons.
In fig. \ref{fig3}, the in-medium excitation energies are shown. 
In the isospin symmetric nuclear matter ($\eta=0$), the considered vectorial ($\o$) interaction is attractive for $\bar{K}\equiv (\bar{K^0}, K^-)$, $D\equiv(D^0, D^+)$ and $\bar{B}\equiv (\bar{B^0}, B^-)$ mesons leading to a drop in the excitation energies of these mesons, whereas it is repulsive for $K \equiv (K^0, K^+)$, $\bar{D}\equiv (\bar{D^0}, D^-)$ and $B \equiv (B^0, B^+)$ mesons in the nuclear medium, leading to an increase in the corresponding excitation energies. The splitting becomes comparatively large in the high-density region.
In isospin asymmetric nuclear matter, the presence of the scalar isovector field $\d$ and the vector isovector field $\rho$ leads to the splitting within the isodoublets. As can be seen from fig. \ref{fig3}(a), in the presence of isospin asymmetry, the $K^0$ excitation energy is seen to increase whereas $K^+$ excitation energy drops in the asymmetric medium compared to its values in SNM. At $\eta=0.5$, $\o_{K^+}$($\o_{K^0}$) is seen to drop(rise) by about $13\%$ and $30\%$ from the values (in MeV) of $545.9(549.6)$ and $698.9(762.4)$ in SNM for densities $\rho_0$ and $3\rho_0$, respectively. This isospin asymmetry effect is observed to be more pronounced at higher baryon densities.
For the $\bar{K}^0$ and $K^-$ mesons, due to the isovectorial ($\d,\rho$) interactions, there is a further drop(rise) in the excitation energy of $\bar{K}^0$($K^-$) from the value in SNM.
For $\eta = 0.5$, $\o_{K^-}(\o_{\bar{K}^0})$ (in $\rm{MeV}$) is observed to be $281.3(422.5)$ and $335.9(-108.5)$ at $\rho_0$ and $3\rho_0$ as compared to the values of $348.9(352.9)$ and $108.2(112.1)$ in SNM. Consequently, the presence of $\d$ and $\rho$ mesons generates a repulsive potential for $K^-$ meson, which indicates a less favorable $K^-$ condensation \cite{tsushima1} in the matter with a higher neutron fraction as compared to the symmetric case. However, the threshold condition $\o_{\bar{K}^0}=0$ implies the possibility of $\bar{K}^0$ condensation in such an environment \cite{MishraK,sarmistha}.
For values of densities of $\rho_0$ and $3\rho_0$, for $\eta=0.5$, the excitation energy $\omega_{D^+}$($\omega_{D^0}$) is observed to decrease(increase) by approximately $4\%$ and $15\%$, respectively from the SNM value. For the density $\rho_0$($3\rho_0$), $\o_{D^-}$ has an increase of around $4\%(11\%)$ and for $\o_{\bar{D}^0}$, an almost similar amount of decrease of around $3.7\%(10.7\%)$, compared to their SNM values. 
The in-medium behaviors of $B(\bar{B})$ mesons are similar to that of $K(\bar{K})$ mesons, owing to the presence of identical light (anti)quark constituents. Figs. \ref{fig3}(e) and \ref{fig3}(f) show the in-medium behavior of the open bottom mesons. In SNM, the excitation energies of the $B$ mesons show an upward trend, whereas the $\bar{B}$ mesons display a decreasing trend.
In asymmetric nuclear matter, as depicted in fig. \ref{fig3}(e), the excitation energies of $B^0$ ($B^+$) experience an increase (decrease) of approximately $1.4\%$ at $\rho_0$, and a rise (drop) of $4\%$ at $3\rho_0$, relative to their values in SNM. On the contrary, $\omega_{B^-}$ and $\omega_{\bar{B}^0}$ are observed to have ascending and descending behaviors respectively, with a deviation of about $1.4\%$ at $\rho_0$, which becomes almost $4.6\%$ at $3\rho_0$ with respect to their values at $\eta=0$.

In the present work, the excitation energies of the pseudoscalar mesons are obtained from the scalar as well as vector interaction terms as represented by eqs. (\ref{44})-(\ref{49}). The excitation energies of the isospin doublet members of $K(\bar{K})$, $D(\bar{D})$ and $B(\bar{B})$ mesons undergo splittings due to isovectorial ($\d,\rho$) interactions in the ANM.
The attractive nature of $D$ and $\bar{B}$ mesons in the nuclear medium can lead to the formation of bound states of these mesons with finite nuclei \cite{tsushimaC}.
The density effects, as well as isospin asymmetry effects on the open strange and open heavy flavor mesons, are observed to be quite appreciable and these can modify the production and propagation of these mesons arising from the asymmetric heavy ion collision experiments.
\section{SUMMARY}
\label{s3}
To summarize we have investigated how the properties of $K\equiv(K^0, K^+)$, $\bar{K}\equiv(\bar{K}^0, K^-)$, $D\equiv(D^0, D^+)$, $\bar{D}\equiv(\bar{D}^0, D^-)$, $B\equiv(B^0, B^+)$ and $\bar{B}\equiv(\bar{B}^0, B^-)$ mesons are affected in the symmetric as well as asymmetric nuclear matter using the QMC$\d$ model. The Quark Meson Coupling model (QMC$\d$), a quark-based approach that describes in-medium hadron properties by considering their internal structure responses to the nuclear medium. Here the medium effects are simulated through the interactions of scalar ($\s$,$\d$) as well as vector ($\o$,$\rho$) meson fields with the confined light quarks (antiquarks) of the hadrons. We have adopted a mean field approximation where the meson fields are replaced by their expectation values. In SNM, there is a drop in the mass of the light quark(antiquark) in the nuclear medium due to the interaction with the scalar $\s$ field. However, the isospin asymmetry effects on the properties of pseudoscalar mesons arise due to the scalar isovector $\d$ and the vector isovector $\rho$ meson. In the presence of the isospin asymmetry in the medium there is a splitting in the masses of quark doublets as well as antiquark doublets due to the interaction with the $\d$ meson. As a consequence, the masses of each isospin doublets of $K(\bar{K})$, $D(\bar{D})$ and $B(\bar{B})$ mesons exhibit splittings in ANM which are observed to increase with increasing baryon density.

In addition, we have studied the excitation energies of the considered pseudoscalar mesons in the nuclear medium. For this, we have adapted a quark-based pseudoscalar meson-nucleon interaction, where the quark and antiquark constituents of the pseudoscalar mesons directly interact with the scalar as well as vector meson fields generated in the nuclear matter leading to the medium modification of their excitation energies. 
Due to the coupling of $\o$ meson, the pseudoscalar mesons along with the light quark(antiquark) are observed to undergo an increase(drop) in their excitation energies, whereas the scalar $\s$ field leads to a drop in their masses as well as excitation energies.
The isospin asymmetry in the medium introduces a splitting of the excitation energies within the members of the isodoublets due to the isovectorial interaction ($\d$ and $\rho$). Whereas, the mass degeneracy is lifted due to the interaction with the $\d$ meson.
In particular, in neutron rich environment, the $\d$ and $\rho$ mesons induce opposite interactions between the mesons within the isospin doublets of $K(\bar{K})$, $D(\bar{D})$ and $B(\bar{B})$ mesons.
The isospin asymmetric effects which are observed to be more pronounced at high densities should have observable consequences in the particle ratios e.g., $K^+/K^0$, $K^-/\bar{K}^0$, $D^+/D^0$, $D^-/\bar{D}^0$, $B^+/B^0$ and $B^-/\bar{B}^0$ in the forthcoming asymmetric heavy ion collisions in Compressed Baryonic Matter (CBM) experiments at FAIR at GSI.
The future experiments at FAIR, with the $\rm{\bar{P}}$ANDA detector at the high-energy storage ring (HESR) \cite{HESR} collider, are poised to provide a comprehensive exploration of the properties of the charm and bottom states, where the present study of open heavy flavor mesons can be relevant.
\begin{section}*{Acknowledgements}
Amruta Mishra acknowledges financial support from the Department of Science and Technology (DST), Government of India (project no. CRG/2018/002226). 
\end{section}

\end{document}